\documentclass{ws-procs975x65}

\begin{document}

\title{Neutrino cosmology - an update}

\author{Steen Hannestad}

\address{Department of Physics, University of Southern Denmark,
Campusvej 55, DK-5230 Odense M, Denmark \\ E-mail:
hannestad@fysik.sdu.dk}


\maketitle

\abstracts{Present cosmological observations yield an upper bound
on the neutrino mass which is significantly stronger than
laboratory bounds. However, the exact value of the cosmological
bound is model dependent and therefore less robust. Here, I review
the current status of cosmological neutrino mass bounds and also
discuss implications for sterile neutrinos and LSND in
particular.}

\section{Introduction}
\label{intro}

The absolute value of neutrino masses are very difficult to
measure experimentally. On the other hand, mass differences
between neutrino mass eigenstates, $(m_1,m_2,m_3)$, can be
measured in neutrino oscillation experiments.

The combination of all currently available data suggests two
important mass differences in the neutrino mass hierarchy. The
solar mass difference of $\delta m_{12}^2 \simeq 7 \times 10^{-5}$
eV$^2$ and the atmospheric mass difference $\delta m_{23}^2 \simeq
2.6 \times 10^{-3}$ eV$^2$
\cite{Maltoni:2003da,Aliani:2003ns,deHolanda:2003nj} (see also the
contribution by C. Giunti to the present volume).

In the simplest case where neutrino masses are hierarchical these
results suggest that $m_1 \sim 0$, $m_2 \sim \delta m_{\rm
solar}$, and $m_3 \sim \delta m_{\rm atmospheric}$. If the
hierarchy is inverted
\cite{Kostelecky:1993dm,Fuller:1995tz,Caldwell:1995vi,Bilenky:1996cb,King:2000ce,He:2002rv}
one instead finds $m_3 \sim 0$, $m_2 \sim \delta m_{\rm
atmospheric}$, and $m_1 \sim \delta m_{\rm atmospheric}$. However,
it is also possible that neutrino masses are degenerate
\cite{Ioannisian:1994nx,Bamert:vc,Mohapatra:1994bg,Minakata:1996vs,%
Vissani:1997pa,Minakata:1997ja,Ellis:1999my,Casas:1999tp,Casas:1999ac,%
Ma:1999xq,Adhikari:2000as},
$m_1 \sim m_2 \sim m_3 \gg \delta m_{\rm atmospheric}$, in which
case oscillation experiments are not useful for determining the
absolute mass scale.

Experiments which rely on kinematical effects of the neutrino mass
offer the strongest probe of this overall mass scale. Tritium
decay measurements have been able to put an upper limit on the
electron neutrino mass of 2.2-2.3 eV (95\% conf.) \cite{Bonn:tw}
(see also the contribution by C. Kraus in the present volume).
However, cosmology at present yields an even stronger limit which
is also based on the kinematics of neutrino mass.

Neutrinos decouple at a temperature of 1-2 MeV in the early
universe, shortly before electron-positron annihilation. Therefore
their temperature is lower than the photon temperature by a factor
$(4/11)^{1/3}$. This again means that the total neutrino number
density is related to the photon number density by
\begin{equation}
n_{\nu} = \frac{9}{11} n_\gamma
\end{equation}

Massive neutrinos with masses $m \gg T_0 \sim 2.4 \times 10^{-4}$
eV are non-relativistic at present and therefore contribute to the
cosmological matter density
\cite{Hannestad:1995rs,Dolgov:1997mb,Mangano:2001iu}
\begin{equation}
\Omega_\nu h^2 = \frac{\sum m_\nu}{92.5 \,\, {\rm eV}},
\end{equation}
calculated for a present day photon temperature $T_0 = 2.728$K.
Here, $\sum m_\nu = m_1+m_2+m_3$. However, because they are so
light these neutrinos free stream on a scale of roughly $k \simeq
0.03 m_{\rm eV} \Omega_m^{1/2} \, h \,\, {\rm Mpc}^{-1}$
\cite{dzs,Doroshkevich:tq,Hu:1997mj}. Below this scale neutrino
perturbations are completely erased and therefore the matter power
spectrum is suppressed, roughly by $\Delta P/P \sim -8
\Omega_\nu/\Omega_m$ \cite{Hu:1997mj}.

This power spectrum suppression allows for a determination of the
neutrino mass from measurements of the matter power spectrum on
large scales. This matter spectrum is related to the galaxy
correlation spectrum measured in large scale structure (LSS)
surveys via the bias parameter, $b^2 \equiv P_g(k)/P_m(k)$. Such
analyses have been performed several times before
\cite{Croft:1999mm,Fukugita:1999as}, most recently using data from
the 2dF galaxy survey \cite{Elgaroy:2002bi}.

However, using large scale structure data alone does not allow for
a precise determination of neutrino masses, because the power
spectrum suppression can also be caused by changes in other
parameters, such as the matter density or the Hubble parameter.

Therefore it is necessary to add information on other parameters
from the cosmic microwave background (CMB). This has been done in
the past \cite{Elgaroy:2002bi,Hannestad:2002xv,Lewis:2002ah},
using ealier CMB data. More recently the precise data from WMAP
\cite{map1} has been used for this purpose
\cite{map2,steen03,el03,Barger:2003vs} to derive a limit of
0.7-1.0 eV for the sum of neutrino masses.

\section{Cosmological data and likelihood analysis}
\label{sec:1}

The extraction of cosmological parameters from cosmological data
is a difficult process since for both CMB and LSS the power
spectra depend on a plethora of different parameters. Furthermore,
since the CMB and matter power spectra depend on many different
parameters one might worry that an analysis which is too
restricted in parameter space could give spuriously strong limits
on a given parameter.

The most recent cosmological data is in excellent agreement with a
flat $\Lambda$CDM model, the only non-standard feature being the
apparently very high optical depth to reionization. Therefore the
natural benchmark against which non-standard neutrino physics can
be tested is a model with the following free parameters:
$\Omega_m$, the matter density, the curvature parameter,
$\Omega_b$, the baryon density, $H_0$, the Hubble parameter,
$n_s$, the scalar spectral index of the primordial fluctuation
spectrum, $\tau$, the optical depth to reionization, $Q$, the
normalization of the CMB power spectrum, $b$, the bias parameter,
and finally the two parameters related to neutrino physics,
$\Omega_\nu h^2$ and $N_\nu$. The analysis can be restricted to
geometrically flat models, i.e.\ $\Omega = \Omega_m +
\Omega_\Lambda = 1$. For the purpose of actual power spectrum
calculations, the CMBFAST package \cite{CMBFAST} can be used.

\subsection{LSS data}

At present there are two large galaxy surveys of comparable size,
the Sloan Digital Sky Survey (SDSS)
\cite{Tegmark:2003uf,Tegmark:2003ud} and the 2dFGRS (2~degree
Field Galaxy Redshift Survey) \cite{2dFGRS}. Once the SDSS is
completed in 2005 it will be significantly larger and more
accurate than the 2dFGRS. At present the two surveys are, however,
comparable in precision and here we discuss constraints from the
2dFGRS alone.

Tegmark, Hamilton and Xu \cite{THX} have calculated a power
spectrum, $P(k)$, from this data, which we use in the present
work. The 2dFGRS data extends to very small scales where there are
large effects of non-linearity. Since we calculate only linear
power spectra, we follow standard procedures and use only data on
scales larger than $k = 0.2\,h~{\rm Mpc}^{-1}$, where effects of
non-linearity should be minimal (see for instance
Ref.~\cite{Tegmark:2003ud} for a discussion). With this cut the
number of data points for the power spectrum reduces to~18.

\subsection{CMB data}

The CMB temperature fluctuations are conveniently described in
terms of the spherical harmonics power spectrum $C_l \equiv
\langle |a_{lm}|^2 \rangle$, where $\frac{\Delta T}{T}
(\theta,\phi) = \sum_{lm} a_{lm}Y_{lm}(\theta,\phi)$. Since
Thomson scattering polarizes light there are also power spectra
coming from the polarization. The polarization can be divided into
a curl-free $(E)$ and a curl $(B)$ component, yielding four
independent power spectra: $C_{T,l}, C_{E,l}, C_{B,l}$ and the
temperature $E$-polarization cross-correlation $C_{TE,l}$.

The WMAP experiment have reported data only on $C_{T,l}$ and
$C_{TE,l}$, as described in Ref.~\cite{map1,map2,map3,map4,map5}

We have performed the likelihood analysis using the prescription
given by the WMAP collaboration which includes the correlation
between different $C_l$'s \cite{map1,map2,map3,map4,map5}.
Foreground contamination has already been subtracted from their
published data.

In parts of the data analysis we also add other CMB data from the
compilation by Wang {\it et al.} \cite{wang3} which includes data
at high $l$. Altogether this data set has 28 data points.

\section{Neutrino mass bounds}
\label{sec:2}

The analysis presented here was originally published in
Ref.~\cite{steen03}, and more details can be found there.

We have calculated $\chi^2$ as a function of neutrino mass while
marginalizing over all other cosmological parameters. This has
been done using the data sets described above. In the first case
we have calculated the constraint using the WMAP $C_{T,l}$
combined with the 2dFGRS data, and in the second case we have
added the polarization measurement from WMAP. Finally we have
added the additional constraint from the HST key project and the
Supernova Cosmology Project. It should also be noted that when
constraining the neutrino mass it has in all cases been assumed
that $N_\nu$ is equal to the standard model value of 3.04. Later
we relax this condition in order to study the LSND bound.

The result is shown in Fig.~1. As can be seen from the figure the
95\% confidence upper limit on the sum of neutrino masses is $\sum
m_\nu \leq 1.01$ eV (95\% conf.) using the case with priors. This
value is completely consistent with what is found in
Ref.~\cite{el03} where simple Gaussian priors from WMAP were added
to the 2dFGRS data analysis. For the three cases studied the upper
limits on $\sum m_\nu$ can be found in Table 1.

%
\begin{table}
\caption{95\% C.L. upper limits on $\sum m_\nu$ for the three
different cases: 1) WMAP+Wang+2dFGRS+HST+SN-Ia, 2)
WMAP+Wang+2dFGRS 3) WMAP+2dFGRS.}
\label{tab:1}       
\begin{center}
\begin{tabular}{cc}
\hline\noalign{\smallskip}
Case &  $\sum m_\nu$ (95\% C.L.) \\
\noalign{\smallskip}\hline\noalign{\smallskip}
1 & 1.01 eV \\
2 & 1.20 eV  \\
3 & 2.12 eV \\
\noalign{\smallskip}\hline
\end{tabular}
\end{center}
\vspace*{1cm}  
\end{table}

In the middle panel of Fig.~1 we show the best fit value of $H_0$
for a given $\Omega_\nu h^2$. It is clear that an increasing value
of $\sum m_\nu$ can be compensated by a decrease in $H_0$. Even
though the data yields a strong constraint on $\Omega_m h^2$ there
is no independent constraint on $\Omega_m$ in itself. Therefore,
an decreasing $H_0$ leads to an increasing $\Omega_m$. This can be
seen in the bottom panel of Fig.~1.

When the HST prior on $H_0$ is relaxed a higher value of $\sum
m_\nu$ is allowed, in the case with only WMAP and 2dFGRS data the
upper bound is $\Omega_\nu h^2 \leq 0.023$ (95\% conf.),
corresponding to a neutrino mass of 0.71 eV for each of the three
neutrinos.

This effect was also found by Elgar{\o}y and Lahav \cite{el03} in
their analysis of the effects of priors on the determination of
$\sum m_\nu$.

However, as can also be seen from the figure, the addition of
high-$l$ CMB data from the Want {\it et al.} compilation also
shrinks the allowed range of $\sum m_\nu$ significantly. The
reason is that there is a significant overlap of the scales probed
by high-$l$ CMB experiments and the 2dFGRS survey. Therefore, even
though we use bias as a free fitting parameter, it is strongly
constrained by the fact that the CMB and 2dFGRS data essentially
cover much of the same range in $k$-space.

It should be noted that Elgar{\o}y and Lahav \cite{el03} find that
bias does not play any role in determining the bound on $\sum
m_\nu$. At first this seems to contradict the discussion here, and
also what was found from a Fisher matrix analysis in
Ref.~\cite{Hannestad:2002xv}. The reason is that in
Ref.~\cite{el03}, redshift distortions are included in the 2dFGRS
data analysis. Given a constraint on the amplitude of fluctuations
from CMB data, and a constraint on $\Omega_m h^2$ , this
effectively constrains the bias parameter. Therefore adding a
further constraint on bias in their analysis does not change the
results.

{\it Neutrinoless double beta decay --} Recently it was claimed
that the Heidelberg-Moscow experiment yields positive evidence for
neutrinoless double beta decay. Such experiments probe the
`effective electron neutrino mass $m_{ee} = |\sum_j U^2_{ej}
m_{\nu_j}|$. Given the uncertainties in the involved nuclear
matrix elements the Heidelberg-Moscow result leads to a mass of
$m_{ee} = 0.3-1.4$ eV. If this is true then the mass eigenstates
are necessarily degenerate, and $\sum m_\nu \simeq 3 m_{ee}$.
Taking the WMAP result of $\sum m_\nu \leq 0.70$ eV at face value
seems to be inconsistent with the Heidelberg-Moscow result
\cite{Pierce:2003uh}. However, already if Ly-$\alpha$ forest data
and a constraint on the bias parameter is not used in the analysis
the upper bound of $\sum m_\nu \leq 1.01$ eV is still consistent.
For this reason it is probably premature to rule out the claimed
evidence for neutrinoless double beta decay.

{\it Evidence for a non-zero neutrino mass --} In a recent paper
\cite{Allen:2003pt} it was noted that there is a preference for a
non-zero neutrino mass if a measurement of the bias parameter from
X-ray clusters is added to the CMB and large scale structure data.
This result arises because the X-ray data prefers a low value of
$\sigma_8$ (bias), which is incompatible with the WMAP and 2dF
result at the 2$\sigma$ level. While this is an interesting
finding it is clear that the X-ray data is subject to a serious
problem with systematic uncertainties, such as the calibration of
the mass-temperature relation. Therefore the result more likely
points to a problem with the interpretation of the X-ray data than
to evidence of a non-zero neutrino mass.

\begin{figure}[h]
\begin{center}
\vspace*{0.0cm} \epsfysize=12truecm\epsfbox{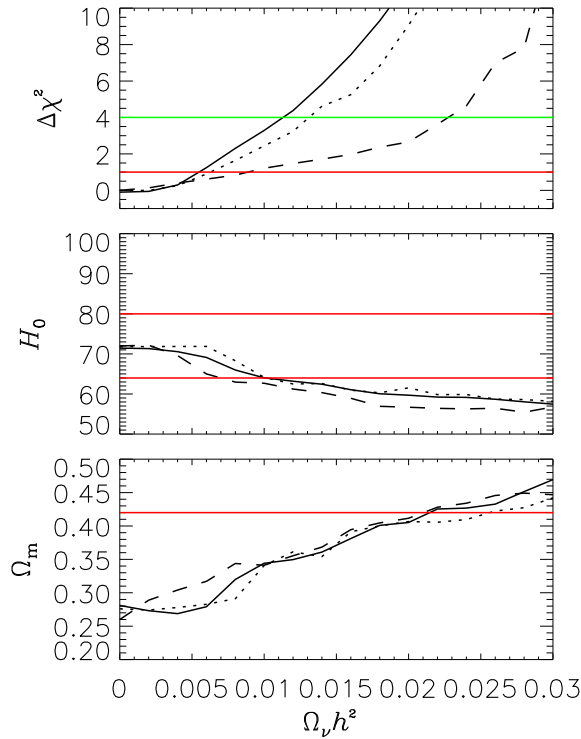}
\end{center}
\vspace*{-1.0cm} \caption{The top panel shows $\chi^2$ as a
function of $\sum m_\nu$ for different choices of priors. The
dotted line is for WMAP + 2dFGRS data alone, the dashed line is
with the additional Wang {\it et al.} data. The full line is for
additional HST and SNI-a priors as discussed in the text. The
horizontal lines show $\Delta \chi^2 = 1$ and 4 respectively. The
middle panel shows the best fit values of $H_0$ for a given $\sum
m_\nu$. The horizontal lines show the HST key project $1\sigma$
limit of $H_0 = 72\pm 8 \,\, {\rm km}\,{\rm s}\,{\rm Mpc}^{-1}$.
Finally, the lower panel shows best fit values of $\Omega_m$. In
this case the horizontal line corresponds to the SNI-a $1\sigma$
upper limit of $\Omega_m < 0.42$.} \label{fig1}
\end{figure}

\section{Sterile neutrinos}
\label{sec:3}

In Ref.~\cite{steen03} it was shown that there is a degeneracy
between the neutrino mass ($\sum m_\nu$) and the relativistic
energy density, parameterized in terms of the effective number of
neutrino species, $N_\nu$.

As can be seen from Fig.~2, the best fit actually is actually
shifted to higher $\sum m_\nu$ when $N_\nu$ increases, and the
conclusion is that a model with high neutrino mass and additional
relativistic energy density can provide acceptable fits to the
data. As a function of $N_\nu$ the upper bound on $\sum m_\nu$ (at
95\% confidence) can be seen in Table 2.

%
\begin{table}
\caption{95\% C.L. upper limits on $\sum m_\nu$ for different
values of $N_\nu$.}
\label{tab:2}       
\begin{center}
\begin{tabular}{cc}
\hline\noalign{\smallskip}
effective $N_\nu$ &  $\sum m_\nu$ (95\% C.L.) \\
\noalign{\smallskip}\hline\noalign{\smallskip}
3 & 1.01 eV \\
4 & 1.38 eV  \\
5 & 2.12 eV \\
\noalign{\smallskip}\hline
\end{tabular}
\end{center}
\vspace*{1cm}  
\end{table}

This has significant implications for attempts to constrain the
LSND experiment using the present cosmological data. Pierce and
Murayama conclude from the present MAP limit that the LSND result
is excluded \cite{Pierce:2003uh} (see also
Ref.~\cite{Giunti:2003cf}).

However, for several reasons this conclusion does not follow
trivially from the present data. In general the three mass
differences implied by Solar, atmospheric and the LSND neutrino
measurements can be arranged into either 2+2 or 3+1 schemes.
Recent analyses \cite{Maltoni:2002xd} of experimental data have
shown that the 2+2 models are ruled out. The 3+1 scheme with a
single massive state, $m_4$, which makes up the LSND mass gap, is
still marginally allowed in a few small windows in the $(\Delta
m^2,\sin^2 2 \theta)$ plane. These gaps are at $(\Delta m^2,\sin^2
2 \theta) \simeq  (0.8 \, {\rm eV}^2, 2 \times 10^{-3}), (1.8 \,
{\rm eV}^2, 8 \times 10^{-4}), (6 \, {\rm eV}^2, 1.5 \times
10^{-3})$ and
 $(10 \, {\rm eV}^2, 1.5 \times 10^{-3})$.
These four windows corresponds to masses of $0.9, 1.4, 2.5$ and
3.2 eV respectively. From the Solar and atmospheric neutrino
results the three light mass eigenstates contribute only about 0.1
eV of mass if they are hierarchical. This means that the sum of
all mass eigenstate is close to $m_4$.

The limit for $N_\nu = 4$ which corresponds roughly to the LSND
scenario is $\sum m_\nu \leq 1.4$ eV, which still leaves the
lowest of the remaining windows. The second window at $m \sim 1.8$
eV is disfavoured by the data, but not at very high significance.

\begin{figure}[h]
\begin{center}
\vspace*{0.0cm} \epsfysize=7truecm\epsfbox{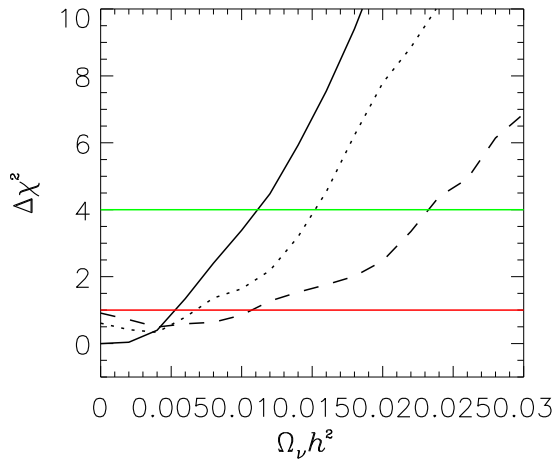}
\end{center}
\vspace*{0.0cm} \caption{$\Delta \chi^2$ as a function of $\sum
m_\nu$ for various different values of $N_\nu$. The full line is
for $N_\nu = 3$, the dotted for $N_\nu = 4$, and the dashed for
$N_\nu = 5$. $\Delta \chi^2$ is calculated relative to the best
fit $N_\nu = 3$ model.} \label{fig2}
\end{figure}

\section{Discussion}
\label{sec:4}

We have calculated improved constraints on neutrino masses and the
cosmological relativistic energy density, using the new WMAP data
together with data from the 2dFGRS galaxy survey.

Using CMB and LSS data together with a prior from the HST key
project on $H_0$ yielded an upper bound of $\sum m_\nu \leq 1.01$
eV (95\% conf.). While this excludes most of the parameter range
suggested by the claimed evidence for neutrinoless double beta
decay in the Heidelberg-Moscow experiment, it seems premature to
rule out this claim based on cosmological observations.

Another issue where the cosmological upper bound on neutrino
masses is very important is for the prospects of directly
measuring neutrino masses in tritium endpoint measurements. The
successor to the Mainz experiment, KATRIN, is designed to measure
an electron neutrino mass of roughly 0.2 eV, or in terms of the
sum of neutrino mass eigenstates, $\sum m_\nu \leq 0.75$ eV (see
contribution by Guido Drexlin to the present volume). The WMAP
result of $\sum m_\nu \leq 0.7$ eV (95\% conf.) already seems to
exclude a positive measurement of mass in KATRIN. However, this
very tight limit depends on priors, as well as Ly-$\alpha$ forest
data, and the more conservative present limit of $\sum m_\nu \leq
1.01$ eV (95\% conf.) does not exclude that KATRIN will detect a
neutrino mass.

Finally, we also found that the neutrino mass bound depends on the
total number of light neutrino species. In scenarios with sterile
neutrinos this is an important factor. For instance in 3+1 models
the mass bound increases from 1.0 eV to 1.4 eV, meaning that the
LSND result is not ruled out by cosmological observations yet.

%
%

\end{document}